\documentclass[journal]{IEEEtran}

\usepackage{graphicx}
\usepackage{color}
\usepackage{multirow}

\usepackage{tabularx}
\usepackage{amsmath}
\usepackage{amsfonts}
\usepackage{amssymb}
\usepackage{float}
\usepackage{xspace}
\usepackage{subcaption}
\usepackage{enumerate}
\usepackage{algorithmicx}
\usepackage{algorithm}
\usepackage{algpseudocode}

\begin{document}
\title {New Normal: Cooperative Paradigm for Covid-19 Timely Detection and Containment using Internet of Things and Deep Learning}


\author{ Farooque Hassan Kumbhar, Syed Ali Hassan, Soo Young Shin ~\IEEEmembership{Fellow,~IEEE}   
\thanks{F. H. Kumbhar, S. A. Hassan and S. Y. Shin are with the IT convergence department at Kumoh National Institute of Technology, Korea. Corresponding author S. Y. Shin, email:wdragon@kumoh.ac.kr}
}
\maketitle

\begin{abstract}
The spread of the novel coronavirus (COVID-19) has caused trillions of dollars in damages to the governments and health authorities by affecting the global economies. The purpose of this study is to introduce a connected smart paradigm that not only detects the possible spread of viruses but also helps to restart businesses/economies, and resume social life. We are proposing a connected Internet of Things ( IoT) based paradigm that makes use of object detection based on convolution neural networks (CNN), smart wearable and connected e-health to avoid current and future outbreaks. First, connected surveillance cameras feed continuous video stream to the server where we detect the inter-object distance to identify any social distancing violations. A violation activates area-based monitoring of active smartphone users and their current state of illness. In case a confirmed patient or a person with high symptoms is present, the system tracks exposed and infected people and appropriate measures are put into actions. We evaluated the proposed scheme for social distancing violation detection using YOLO (you only look once) v2 and v3, and for infection spread tracing using Python simulation. 
\end{abstract}


\IEEEpeerreviewmaketitle

\section{Introduction}
\label{intro}
The coronavirus (COVID-19) is a contagious virus from the severe acute respiratory syndrome (SARS) family which affects the respiratory system of the host with high fever, cough, and breathing problems. The virus has spread to 213 countries with devastating effects, starting in 2019-Dec with more than 13.4 million reported cases and 0.58 million deaths by 15 July 2020\footnote{Total cases: 13.4 million, death: 581,221, active cases: 4.9 million, highest daily cases: 236,000 on 10-July-2020 [Online: https://www.worldometers.info/coronavirus/]}. The contagious virus spread by direct exposure to the infected hosts (human, animals) through a cough or sneeze droplets which can travel up to 2 meters (6 feet). The risk of getting infected increases with a close encounter with the confirmed patient without any intermediate preventive coverings such as face masks, glass shields, eye protectors, etc~\cite{Chamola_ia2020}. Various regular business and social activities such as having coffee at a coffee shop, getting haircuts from barbers, walking in a group, taking public transports, etc. lead to human-to-human infection spread. Various governments took extreme measures of countrywide lockdown and time-based curfews to reduce the number of new cases. These steps however cost too much in terms of business, government financial support, and personal psychological impacts. Stopping the spread of COVID-19 is essential but resuming daily life activities is also important. Standard operating procedures (SOPs) with smart lockdowns to minimize the possible spread of infection can help achieve the goal of restoring economies in a safe environment.

\begin{figure*}[t]
	\centering
	\includegraphics[width=0.8\textwidth]{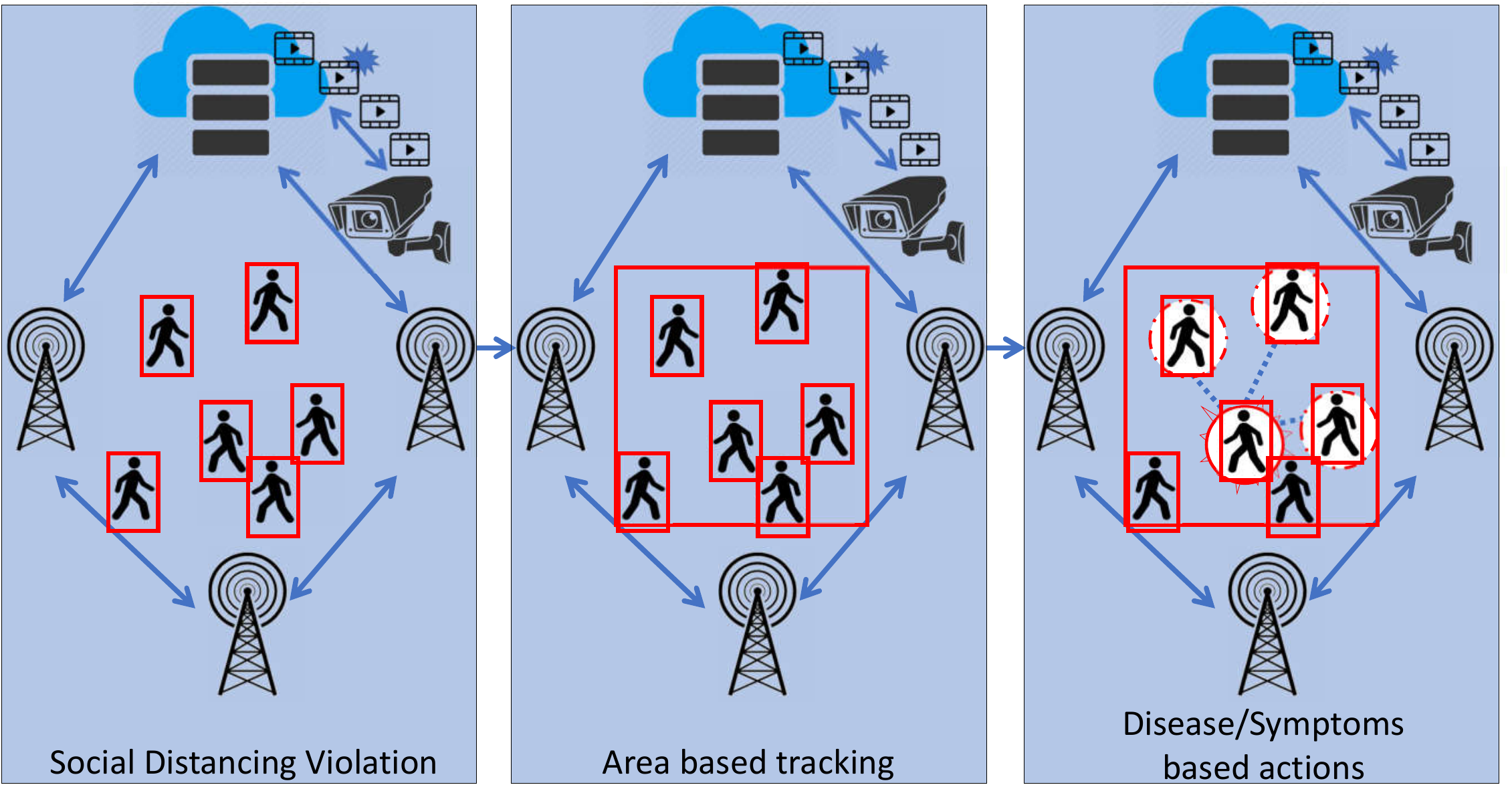}
	\caption{Proposed IoT and CNN based timely detection scheme, spanning over three major phases}
	\label{fig_proposed}
\end{figure*}

To open businesses and economies, it is essential to conduct mass screening~\cite{Hossain_in2020}, infection tracing, confirmed patient monitoring, and timely detection to reduce further spread by taking appropriate actions. We believe that the Internet of Things (IoTs) with smart wearable, fog computing~\cite{Mobasheri_iotm2020}, and the smart connected city has a lot to offer~\cite{Awaisi_iotm2020}. The contact tracing using mobile applications is already being used by various countries such as Australia, Singapore, and South Korea to track people with the confirmed disease and newcomers in self isolation~\cite{Abbas_icem2020}~\cite{Hernandez_ia2020}. Google+Apple and Facebook have stepped up to provide their service for contact tracing as well~\cite{Michael_icem2020}. A partial solution in~\cite{Stojanovic_meco2020} proposes a headset like wearable device which can track COVID-19 symptoms. Authors in~\cite{Punn_ax2020} propose a surveillance system to monitor social distancing between a group of people using convolution neural network based object detection technique; YOLO (you only look once) v3~\cite{Redmon_ax2018}. An interesting study in~\cite{Alsaeedy_iomb2020} proposed to identify regions with high mobility using cellular handovers by mapping the relationship of cellular mobility to disease spread. However, researchers are working tirelessly to combat the challenges of the COVID-19 pandemic. However, we believe that there is a dire need of a connected paradigm which not only restricts further virus spread but also allows people to resume their life.

In this study, we propose a connected IoT-based paradigm that targets two major agendas; 1) timely detection and appropriate actions to stop the spread and 2 ) new normal with connected and informed resumption of daily life activities. First, a connected surveillance camera detects violations of social distances by detecting objects and reporting for the possible spread of viruses~\cite{Ali_ictc19}. Thereafter, the fog node-based server~\cite{Antonini_iotm2019} traces connected cellular devices for active confirmed patients in the reported area. Each wearable device based on IoT assists in identifying people with high symptoms in the reported area. If a confirmed patient or a person with high symptoms is in the reported area, the system trigger warnings and/or actions to contain the spread of the virus, using the connected health care system~\cite{Guo_itii2020}. Not only does the proposed paradigm restricts the spread of the virus, but it also allows a more secure and informed environment to restart new normal. Our major notable contributions are:   
\begin{enumerate}
	\item Detection of social distance breaches using YOLO v2 and YOLO v3 based on CNN.
	\item Area-based tracking of cellular user activity.
	\item Detection of active diseased persons in the area.
	\item Detection of a highly symptomatic person using a smart wearable.
	\item Contact tracing of persons in the reported area at risk (confirmed illness or high symptom).
	\item Timely actions and warnings (quarantine or self-isolation) toward persons at risk.
\end{enumerate}

\section{Proposed New Normal Solution}
\label{proposed}
Our proposed scheme has three major phases, as illustrated in Fig.~\ref{fig_proposed}, in which we detect social distance, area of risk, and personal (exposed or infected) tracing.
\subsection{Social Distancing Violation Detection}
\begin{figure}
	\centering
	\includegraphics[width=\columnwidth, height=100mm]{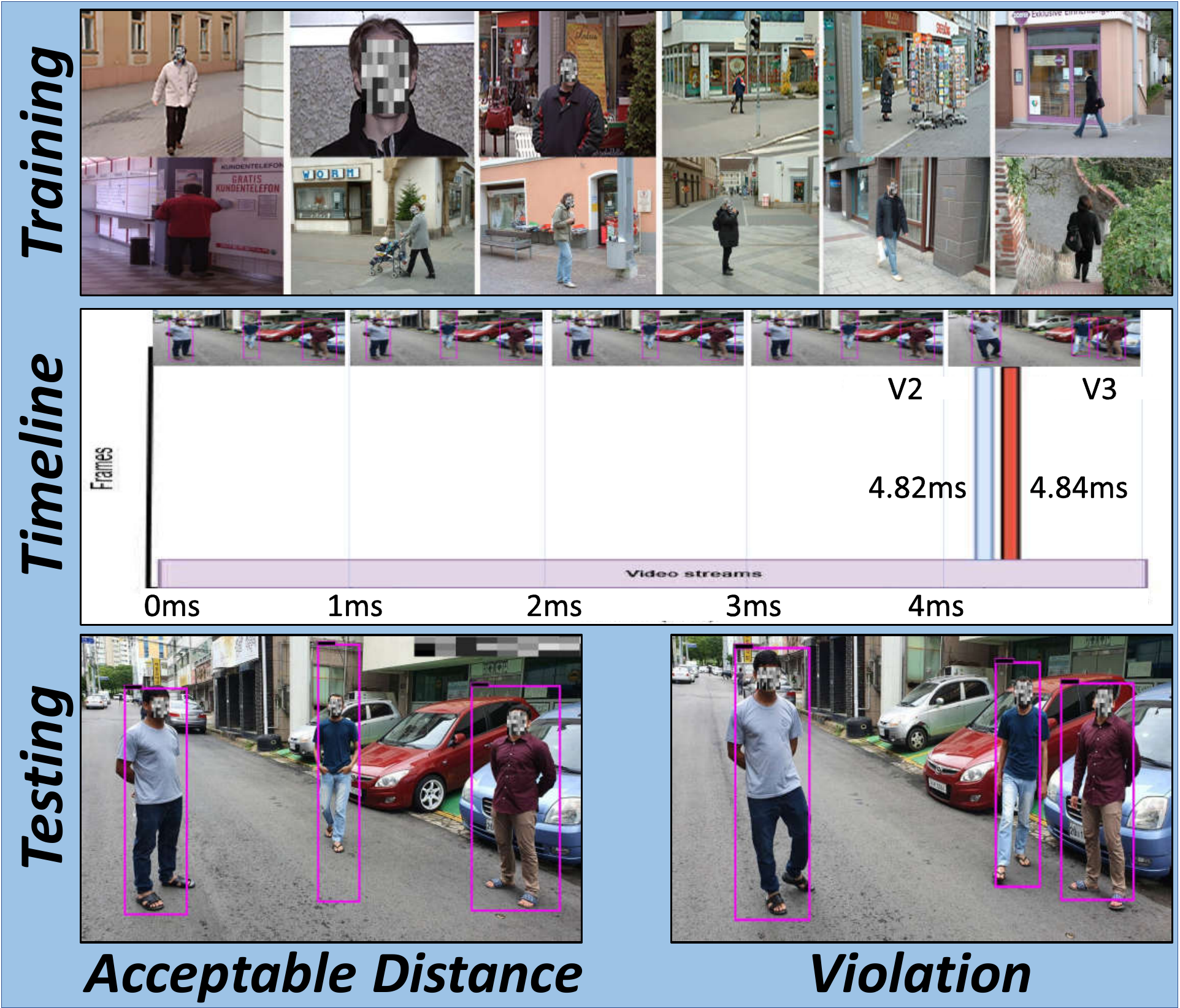}
	\caption{Social distance violation detection using CNN based object detection (YOLOv2 and YOLOv3)}
	\label{fig_pictures}
\end{figure}

The continuous video stream by surveillance cameras is fed to fog node based object detection model. Each frame of the video is considered as a separate image and a CNN based object detection assigns boundary boxes to each object in the picture. The CNN based model is trained with a number of desired objects for a single class detection, in this case, person. Fig.~\ref{fig_pictures} shows the training images which consist of random people so that the model can detect any person. Our training has an image database of 2,000 pictures, marked with person class. We considered two methods of object detection which are widely used, YOLO v2 and YOLO v3~\cite{Fang_ia2020}. YOLO models accurately and quickly detect objects classified by multiple labels by applying a single neural network to the image. The image is divided into several regions where each detected object is assigned a bounding box with features, such as center coordinates, height and width, confidence, and object class shape accuracy. The YOLO v3 is an improved incremental approach (over YOLO v2) with 53 convolutional layers and a deep network darknet-53 backbone. Given a test image, the CNN based model produces a collection of all the detected objects. It can be seen in Fig.~\ref{fig_pictures} that given a video stream over a timeline, the object detection method identifies the violation immediately. An example of testing images is also depicted where the inter-object distance is acceptable or it is in the violation. The violation triggers area and disease based tracing for appropriate actions. 

\begin{algorithm} [t]
	\begin{algorithmic}[1]
		\caption{Proposed New Normal Algorithm}
		\State Camera share continuous video stream with the server
		\ForAll {Video frames}
			\State Detects objects and assigns bounding boxes using CNN based object detection (YOLO v2 or YOLO v3)
			\ForAll {Pair of detected objects}
				\If {Exists $D_{ij}<D_{th}$}
					\State Server identifies area by communicating with BSs (associated with camera)
					\ForAll{Active users in BS(s) area}
						\State Active user health update using health center information 
						\If {Exists a user with confirmed disease}
							\State Immediate health and safety actions
						\Else
							\State Active user health vitals using IoT based smart wearable or last handheld checkup
							\If{Exists a user with high symptoms}
								\State Immediate notification for self isolation to affected users based on proximity
							\EndIf 
						\EndIf
					\EndFor
				\EndIf
			\EndFor
		\EndFor
		
		 where $D_{ij}$ is the inter-object distance between boundary boxes of detected objects, $D_{th}$ is the allowed minimum distance threshold, 
		\label{algo_newNormal}
	\end{algorithmic}
\end{algorithm}

\begin{table}[] 
	\centering
	\caption{Simulation Parameters}
	\label{tab_par}
	\begin{tabular}{|p{1.9cm}|p{1.9cm}|p{1.9cm}|p{1.1cm}|}
		\hline
		\multicolumn{4}{|p{8cm}|}{\textbf{\textit{Simulation parameters CNN based object detection (YOLO v2 and YOLO v3 training parameters)}}}\\
		\hline
		\textbf{Parameters}   & \textbf{Values}      &  \textbf{Parameters}   & \textbf{Values}   \\ \hline
		Optimizer              & SGD (Stochastic Gradient Descent)~\cite{Fang_ia2020}        &  Iterations              & 10,000~\cite{Fang_ia2020}\\ \hline
		Input image dimension       & 416 X 416~\cite{Fang_ia2020}        &                 Exposure           & 1.5~\cite{Fang_ia2020}  \\ \hline
		Learning rate                       & 0.001~\cite{Fang_ia2020}                &   Channels          & 3~\cite{Fang_ia2020}             \\ \hline
		Batch size   & 64~\cite{Fang_ia2020}         &   Decay          & 0.0005~\cite{Fang_ia2020}   \\ \hline
		Subdivisions   & 4~\cite{Fang_ia2020}      &   Momentum           & 0.9~\cite{Fang_ia2020}         \\ \hline
		Stride           & 1~\cite{Fang_ia2020}     &   Hue           & .1~\cite{Fang_ia2020}                                              
		\\ \hline
		Saturation           & 1.5~\cite{Fang_ia2020}       &               & 	\\ 
		\hline 
		\multicolumn{4}{|p{8cm}|}{\textbf{\textit{Simulation parameters python based virus spread simulation}}}\\
		\hline 
		Time &	300 min	& Active confirmed patient &	50\\ 
		\hline 
		People count	&150 to 2500	& R0 without Masks & 	0.6\\ 
		\hline 
		Area	& 1,000 x 1,000 meters	& R0 with Masks 	& 0.3\\ 
		\hline 
		Mobility	& Random Walk	& Contact duration with confirmed patient 	& 10 m \\ 
		\hline 
		Mobility Speed	& [1, 10] meter / min	& Symptoms persistence duration 	& 60 m\\ 
		\hline 
		Distance Threshold	& 5 meters 	& Symptom Threshold	& 0.9\\ 
		\hline 
	\end{tabular}
\end{table}

\subsection{Area at Risk Tracking and Disease/ Symptoms Based Actions}
The connected cellular devices in the area at risk are identified using cellular data. Each camera feed is also mapped to the cellular base stations (BSs) in that area.  The active mobile users are scrutinized on two aspects: 
\begin{enumerate}
	\item Confirmed disease cases: We assume that the health centers have data of confirmed active cases. If a cellular user in the reported area is a diagnosed active case then the health officials are notified. Notification to the active users in the area informs about a possible exposure which also requires them to self-isolate or to contact a medical center. Moreover, the active confirmed patient is immediately quarantined. 
	\item People having high symptoms for COVID-19: The people who have been in contact with active confirmed patients or have been traveling might not show symptoms immediately. We consider that each person is either wearing an IoT based health monitor or getting health vitals checked frequently using handheld equipment. Each disease, specially COVID-19 has widely known symptoms that can help in identifying a possible infection. The IoT based smart wearable can monitor major observed symptoms for COVID-19, such as fever (98.6$\%$ cases), fatigue (70$\%$ cases), and dry cough (60$\%$ cases). A person having symptoms higher than a specified threshold for a minimum specific time is also a major concern. If a person with high symptoms is recorded regardless of the reported area, then immediate notifications are dispatched. The person and the people in the close proximity are promptly asked to self-isolate.
\end{enumerate}
It should be noted that the area that has no violations, confirmed patients or a person with high symptoms, is not at risk and can resume activities in the social life.

\subsection{Proposed Connected Paradigm}
Algorithm~\ref{algo_newNormal} outlines the proposed new normal algorithm where surveillance cameras continuously monitor to detect any social distancing violations. Each video frame is passed to the pre-trained CNN based image detection model. The CNN based model returns all the detected objects (person class) in the given picture with boundary boxes. Fig~\ref{fig_pictures} shows the test images, where three objects are detected and assigned boundary boxes. For every pair of detected objects, an inter-object distance is compared with a pre-defined and programmable distance threshold. If the objects (people class) violate the guidelines for social distancing then the system will request all active BSs to track active users. With the information available at health centers, it is identified that any active user in that area has been confirmed for the infectious disease or that any user has persistent high symptoms. If there is any confirmed patient present in the area then immediate action is taken by health authorities. On the other hand, if a person has high symptoms, he/she is notified to take immediate self-isolation precautionary and to notify health officials. The people in the area at risk are notified without revealing the identities, to take precautions, and contact health officials if feeling sick. 

Our proposed scheme allows people to perform day-to-day business and operations using smart wearables to keep track of documented patients and potential patients. We believe that the proposed system creates a new normal environment, which can help reopen economies and reduce the wide spread of the disease.

\begin{figure}
	\centering
	\includegraphics[width=0.8\columnwidth]{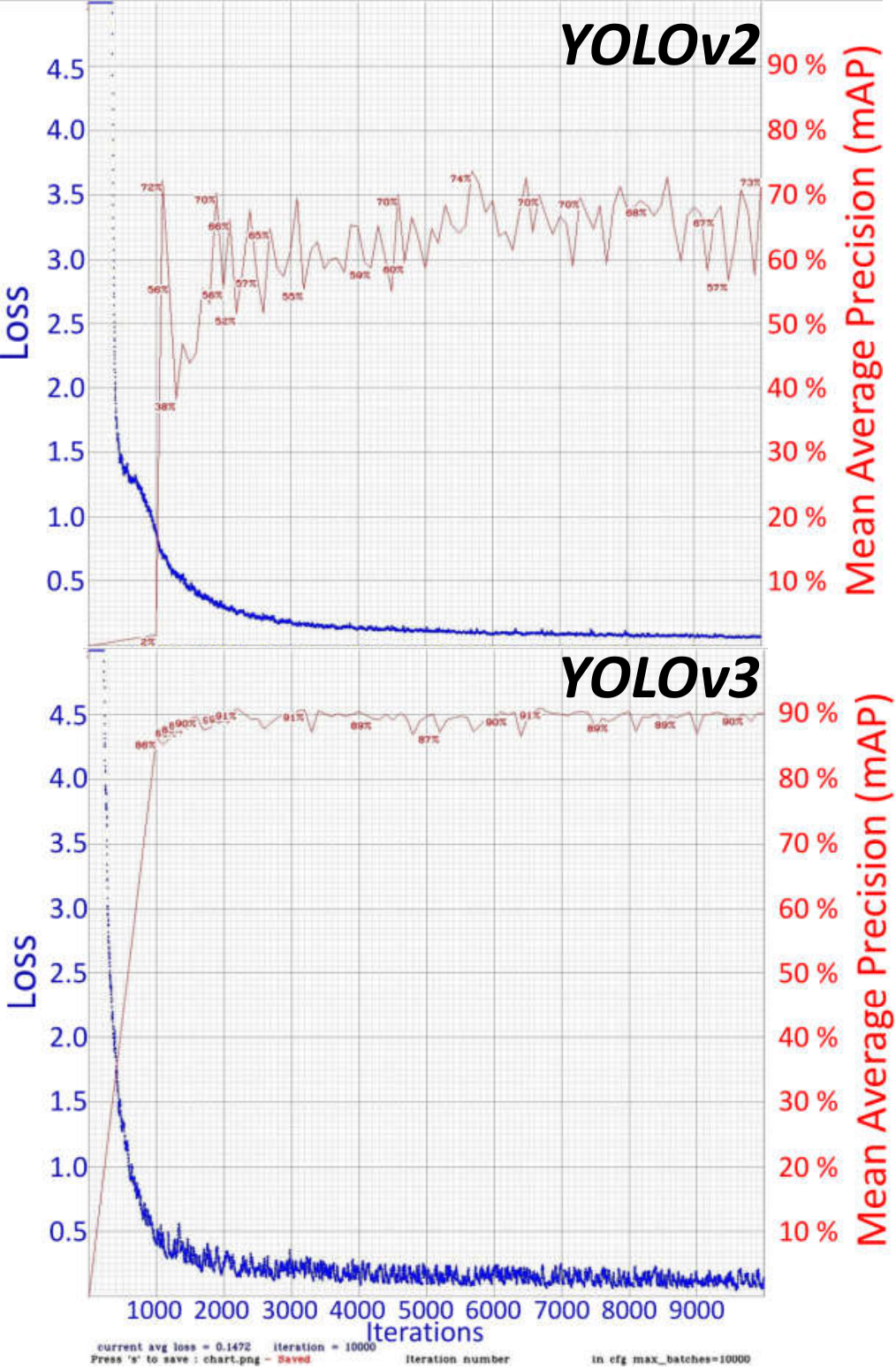}
	\caption{Object detection mAP for YOLO v2 and YOLO v3}
	\label{fig_mAPresults}
\end{figure}

\section{Experiments and Performance Evaluation}
\label{Eval}
We evaluate the proposed scheme in two perspectives; 1) Object detection using CNN (YOLO v2 and YOLO v3), and 2) Virus spread using python simulation. 

\textbf{CNN based object detection:} 
We have evaluated two widely used CNN based object detection algorithms; YOLO v2 and YOLO v3~\cite{Redmon_ax2018} using a system with GPU NVIDIA Titan. The darknet-19 for YOLO v2 and darknet-53 for YOLO v3 are trained on our dataset~\cite{Fang_ia2020}. The model is trained with 2,000 images each having a resolution of 416 $\times$ 416, with a batch size of 64 and the subdivision of 8. The learning rate of 0.001 ensures faster model convergence with early training time~\cite{Fang_ia2020}. We ran 10,000 iterations with Stochastic Gradient Descent (SGD) optimizer. A list of all training parameters is outlined in Table~\ref{tab_par}. The mean average precision (mAP) is calculated with the train/test split set of 80/20 ratio. Fig.~\ref{fig_mAPresults} illustrates the comparative results generated by the trained CNN based models of YOLO v2 and YOLO v3. The results smooth in both the models after 3,000 iterations with very small loss value. However, the YOLO v3 outperforms YOLO v2 by achieving the mAP value of up to 90 $\%$ as opposed to mAP value of 73 $\%$.

\begin{figure}
	\centering
	\includegraphics[width=0.8\columnwidth]{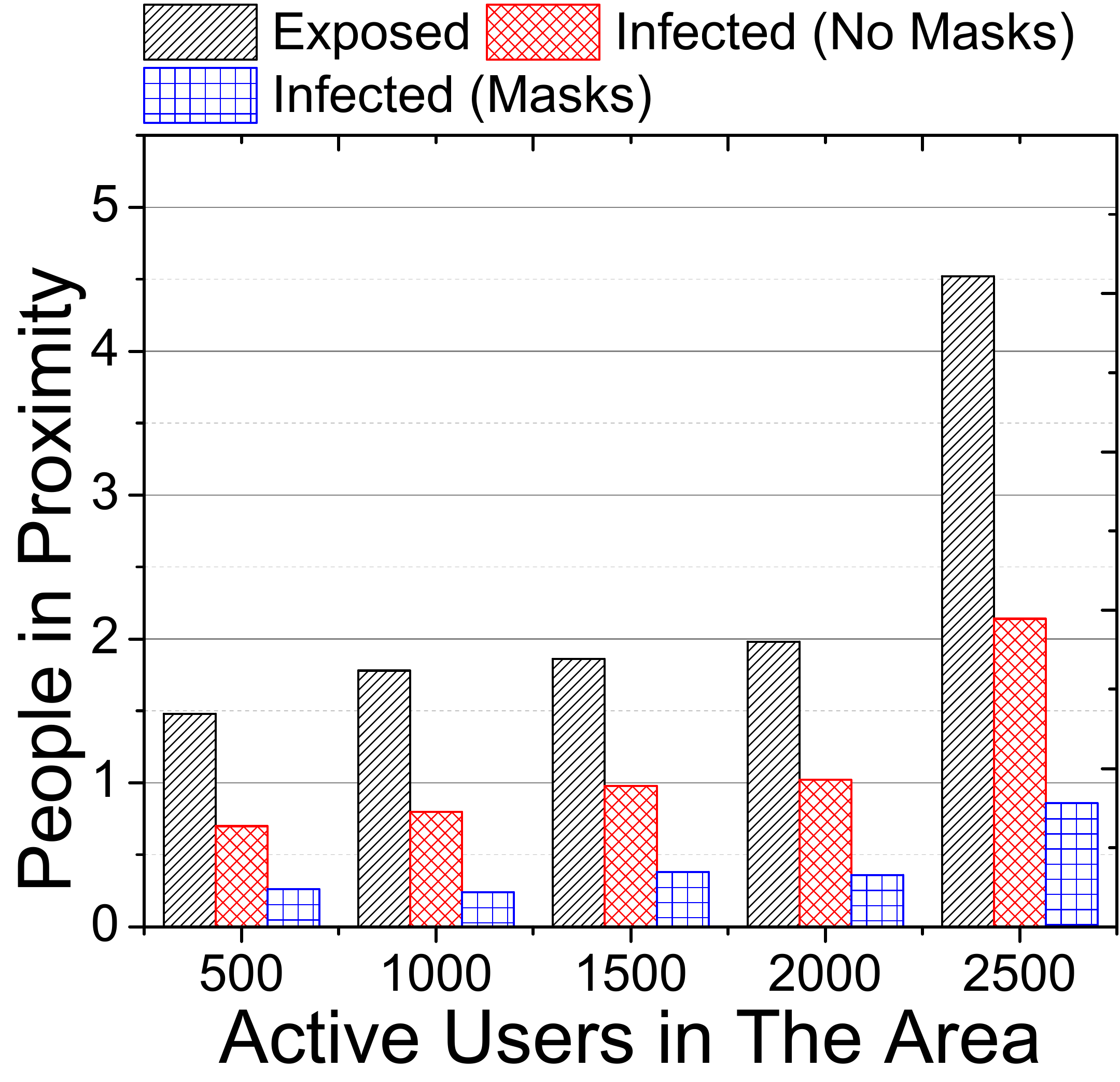}
	\caption{People in the proximity of the people with confirmed disease}
	\label{fig_diseased}
\end{figure}

\begin{figure}
	\centering
	\includegraphics[width=0.8\columnwidth]{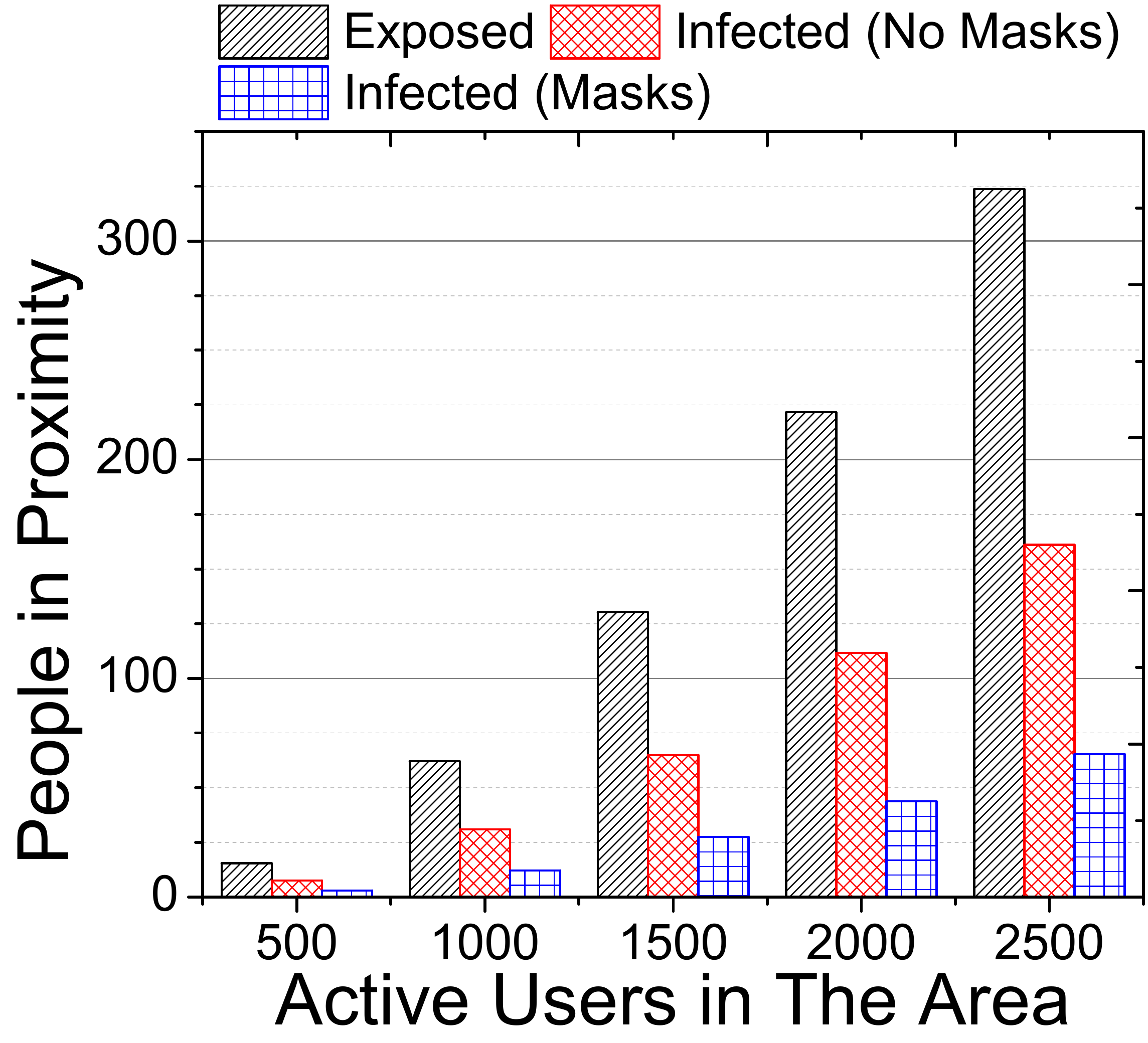}
	\caption{People in the proximity of a person having high symptoms (potential virus carriers)}
	\label{fig_symptoms}
\end{figure}

\textbf{Python simulation for virus spread:}
Considering that the CNN based object detection module generates a violation warning in a particular area of 2,000 $\times$ 2,000 meters having 500 to 2,500 active users. Each user depicts random mobility between [0, 20] meters per minute. Our python based simulation assigns a wearable health monitoring device to each active user which continuously reports health vitals for an observation period of 500 minutes. We plant 50 confirmed patients on random locations in the reported area. A user is exposed to the disease by being in the proximity (distance threshold 3 meters) of a confirmed patient or a person with high symptoms. However, the probability of getting infected if everyone is wearing masks or not wearing masks is set to an arbitrary programmable value of 0.5 and 0.2, respectively. Moreover, the symptoms based detection identifies a possible infected person if he/ she has symptoms higher than a predefined threshold (0.9), for an hour duration. 

Fig.~\ref{fig_diseased} illustrates the potentially infected count of exposed persons to the people having confirmed disease. With the increase in the number of people from 500 to 2500, the risk of getting exposed also heightens. However, wearing protective gear like a face mask or eye masks substantially reduces the probability of getting infected by 50$\%$. On the other hand, Fig.~\ref{fig_symptoms} shows that the number of people exposed to a person having high symptoms (potential carrier) increase from 20 to 350+ with the increase in the total active users from 500 to 2500, respectively. The protective gears reduce the risk of getting infected after having exposure by more than 50$\%$. Nevertheless, the proposed scheme's timely detection of exposed or infected help in taking appropriate actions. Only the detected people are required to be isolated or tested, whereas others carry on with the social lives in the new normal.

\section{Conclusion}
\label{concl}
In this study, we propose a connected paradigm by using IoT based health monitoring and CNN based object detection methods. The proposed scheme aims to contain the spread as soon as possible and allow people to continue with their social activities. Our scheme identifies the social distancing violations using CNN based object detection and tracks exposed or infected people using a smart wearable. The YOLO v3 darknet-53 model based on CNN achieves 90$\% $accuracy in object detection to identify inter-object distance and violation of social distancing. In addition, the python simulation successfully traces all exposed people with the likelihood that they will get infected with and without masks. We believe that this pandemic needs to be evaded by the proposed connected paradigm and can be a fundamental system for future disasters.

\section*{Acknowledgment}
This work has been supported by the National Research Foundation of Korea (NRF) with grant no. 2019H1D3A1A01102978.

\begin{IEEEbiography}[{\includegraphics[width=1in,height=1.25in,clip,keepaspectratio]{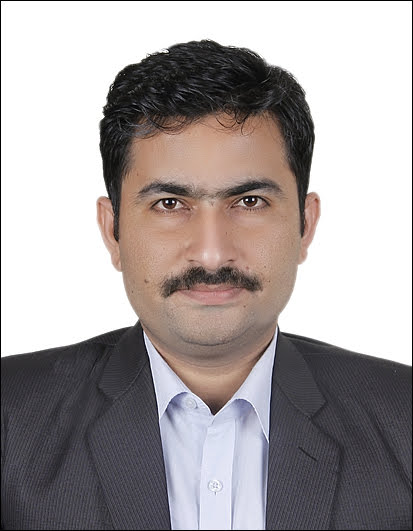}}]
	{Farooque Hassan Kumbhar} is a postdoctoral researcher at the School of Electronics, Kumoh National Institute of Technology, Gumi, South Korea. Between 2017 and 2020 he was an assistant professor in the Department of Computer Science, National University of Computer and Emerging Sciences, Pakistan. In 2017, he received his MS and Ph.D. degree from the College of Software, Sungkyunkwan University, South Korea. His research findings cover privacy, mobile communications, C-RAN, Internet of Things, and Machine to Machine Communication. His work is well published with IEEE, Springers, IETE, and IET. He is also director of the research group named as intelliNet: http://intellinet.rf.gd/ Email: farooque.hassan@nu.edu.pk
\end{IEEEbiography}

\begin{IEEEbiography}[{\includegraphics[width=1in,height=1.25in,clip,keepaspectratio]{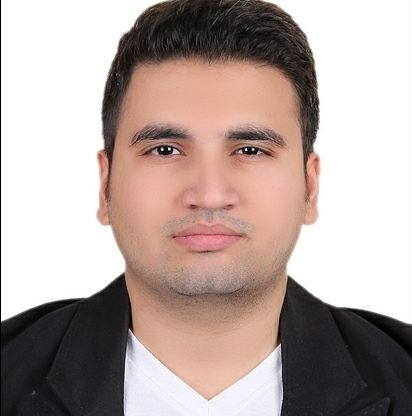}}]{Syed Ali Hassan} completed his master's degree from Kumoh National Institute of Technology, Gumi, South Korea and currently working in korea as an AI/AR Application Software Developer. He completed his B.S. ('16) from the Department of Software Engineering, Mohammad Ali Jinnah University, Pakistan. His interests are in Mixed Reality, Augmented Reality, Virtual Reality, and Deep Learning. Email: syedali1621@gmail.com
\end{IEEEbiography}

\begin{IEEEbiography}[{\includegraphics[width=1in,height=1.25in,clip,keepaspectratio]{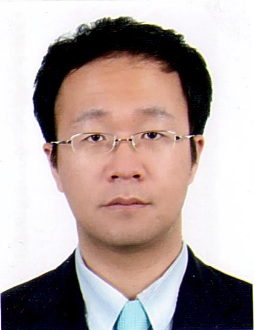}}]{Soo Young Shin}  (M’07–SM’17) received his Ph.D. degrees in electrical engineering and computer science from Seoul National University in 2006. He was with WiMAX Design Lab, Samsung Electronics, Suwon, South Korea from 2007 to 2010. He joined as a full-time professor at School of Electronics, Kumoh National Institute of Technology, Gumi, South Korea from 2010. He is currently an Associate Professor. He was a post-Doc. researcher at the University of Washington, USA in 2007. In addition, he was a visiting scholar to University of the British Columbia, Canada in 2017. His research interests include 5G/6G wireless communications and networks, signal processing, Internet of things, mixed reality, drone applications, etc. Email: wdragon@kumoh.ac.kr
	
\end{IEEEbiography}

\end{document}